\documentclass[aip,pop,floatfix]{revtex4-1}
\usepackage[]{algorithm2e}
\usepackage{hyperref}
\usepackage{amsmath}
\usepackage{amsfonts}
\usepackage{amssymb}
\usepackage{graphicx}
\usepackage{subfigure}
\usepackage{verbatim}
\usepackage{bm}
\usepackage{rotating}
\usepackage{multirow}

\begin{document}

\title{Stochastic simulation of collisions using high-order weak convergence algorithms}

\author{Wentao Wu}

\affiliation{School of Physics, University of Science and Technology of China,
Hefei, Anhui 230026, China}

\author{Jian Liu}

\affiliation{School of Physics, University of Science and Technology of China,
Hefei, Anhui 230026, China}

\thanks{Corresponding author, jliuphy@ustc.edu.cn}

\author{Hong Qin}

\affiliation{School of Physics, University of Science and Technology of China,
Hefei, Anhui 230026, China}

\affiliation{Plasma Physics Laboratory, Princeton University, Princeton, NJ 08543,
USA}

\begin{abstract}
Electron collisions, described by stochastic differential equations (SDEs), 
were simulated using a second-order weak convergence algorithm. 
Using stochastic analysis, we constructed an SDE for energetic electrons in 
Lorentz plasma to incorporate the collisions with heavy static ions
as well as background thermal electrons. 
Instead of focusing on the errors of each trajectory, 
we initiated from the weak convergence condition to directly improve 
the numerical accuracy of the distribution and its moments. 
A second-order weak convergence algorithm was developed, 
estimated the order of the algorithm in a collision simulation case, 
and compared it with the results obtained using 
the Euler-Maruyama method and the Cadjan-Ivanov method. For a fixed error level,
the second order method largely reduce the computation time compared to
the Euler-Maruyama method and the Cadjan-Ivnaov method.
Finally, the backward runaway phenomenon was numerically studied and 
the probability of this runaway was calculated using the new method.
\end{abstract}

\maketitle

\section{Introduction}
Monte Carlo method is a approach that calculates the distribution or expectation
through random sampling and simulations.
It is commonly used to simulate collision in plasma
by first solving the corresponding Newton equation with random forces 
and then statistically reconstructing the distribution function. 
Random force was originally formulated by Paul Langevin\cite{Langevin1908} 
and the equation was named after him. It was proposed as an intuitive approach 
to explain Brownian motion after diffusion theory was developed by Einstein\cite{Einstein1906}. 
A rigorous mathematical framework for this remained undiscovered 
until Wiener and Ito formulated the integral theories of the Wiener process 
and the stochastic differential equation (SDE) theory\cite{Ito1974,Klebaner2005,Oeksendal2003,Karatzas1991,higham2001algorithmic}. 
According to their works, the effect of the random force was 
interpreted as an integral in a Wiener process rather than 
a simple random variable plugged into an ordinary differential equation. 
Thus, the Langevin equation was reformed as an SDE and its connection to 
diffusion theory was revealed by the famous Feynman-Kac formula, 
which boosted the research on SDEs. 
In most applications\cite{Albright2002,Albright2003,Cadjan1999,Cadjan1997,Castejon2003,Fernandez2012,Manheimer1997,Rosin2014,shkarofsky1992numerical}, 
the SDE is solved with numerical algorithms. Various numerical algorithms 
have been proposed to solve this equation. 
Most studies simply adopted the Euler-Maruyama method\cite{Jones1996,Manheimer1997,Albright2002,Albright2003}, 
and some used the Milstein method\cite{mil1975approximate}. 

Specific to the research of collision in plasma physics, 
SDE have mainly been used to study the collision effects. 
By applying the traditional Langevin approach, 
Jones and Manheimer\cite{Jones1996,Manheimer1997} separately developed 
Coulomb collision models in particle-in-cell (PIC) simulations. 
Cadjan and Ivanov\cite{Cadjan1999,Cadjan1997} first expressed 
the Lorentz-collision operator in a modern SDE form. 
Subsequently, Albright developed the quiet direct simulation 
Monte Carlo (QDSMC)\cite{Albright2002,Albright2003} 
technique utilizing the Ito stochastic integral. 
These methods have been applied to the study of wave-particle 
interactions\cite{Castejon2003} and runaway electrons\cite{Fisch1987,Karney1986}. 
Additionally, Cadjan and Ivanov proposed a numerical method\cite{Cadjan1999,Cadjan1997}, 
which is referred to as the Cadjan-Ivanov method in this study.

However, the numerical errors in traditional procedures for 
solving the Langevin equations have not been carefully discussed. 
Traditional improvement of numerical schemes mainly focused on 
finding a correct and accurate path. 
As a result, the measurement of the accuracy of a numerical solution 
was judged by its deviation from theoretical paths. 
This criterion can be strictly explained using the mathematical concept of strong convergence. 
While, for Monte Carlo simulation of collisions, 
it is more important to pursue an accurate distribution function or 
its moments instead of the path itself. 
In other words, although the strong convergence condition 
implies the correctness of the distribution function, 
it need not be necessarily met as the condition is too strong. 
Consequently, the viability of a numerical algorithm should be evaluated 
based on the deviation of the numerical distribution from the theoretical distribution, 
which is referred to as the weak convergence condition. 

The order of both the Euler-Maruyama method and the Cadjan-Ivanov method 
was 0.5 under strong convergence. 
The path of a numerical solution generated by 
these traditional methods converges to an exact solution with a numerical error 
on the order of $\sqrt{\mathrm{\Delta }}$. From the viewpoint of weak convergence, 
the distribution of numerical solutions of the two methods is convergent 
to the distribution of exact solutions on the order of one. 
Occasionally, even the strong convergence order has been improved to 1.0, 
as with the Milstein method\cite{Kloeden1992,Milstein2004}, 
while the weak convergence order remained at this value. More precisely, 
the accuracy of the distribution function does not improve 
even if the path of a particle is calculated with a more accurate algorithm. 
Therefore, to improve the accuracy of an Monte Carlo simulation, 
increasing the convergence rate from a weak convergence aspect 
is more efficient and economic. 

We developed a second order Monte Carlo simulation algorithm using modern SDE 
framework under weak convergence for extended Lorentz collision operator, 
which not only includes the collision effects of energetic electron with 
background ions yielded the standard Lorentz operator, 
but also the collision with background electrons. 
The SDE of the energetic electrons are derived from the Boltzmann equation 
for such electrons with extended Lorentz collision terms using the 
Dynkin formula\cite{Klebaner2005,Karatzas1991}. By further assuming the 
diffusion coefficient matrix to be symmetric, an analytical SDE form can be 
derived using Cadjan and Ivanov's decomposition method\cite{Cadjan1997}. 
By applying the Ito-Taylor expansion, the solution of the Ito SDE can be 
expanded into the sum of a series of Ito multiple integrals\cite{Kloeden1992}. 
Under weak convergence conditions, these integrals can be further simplified by 
replacing the distribution of the increment of the Wiener process-which is 
a normal distribution-with a three-point distribution\cite{kloeden1992higher,Kloeden1992}. 
By dropping the remainder terms, a second-order algorithm under 
the weak convergence condition can be derived. 

To test the performance of the second-order method, the numerical order of 
the algorithm was estimated in a backward runaway\cite{Karney1986} simulation
and compared with those estimated using the Euler-Maruyama and the Cadjan-Ivanov methods. 
To estimate the error of each method and time steps, 
a good enough solution was treat as the theoretical value.
The good enough solution was required to meet two criterions.
First, the solution should be close to that of nearest larger time steps using the same method,
which is tested using a Welch's t-test\cite{welch1947}.
Second, the solution should be close to that of different methods using the same time step,
which is tested usign a ANOVA F-test\cite{fisher1936use}.
Numerical orders of algorithms were estimated using 
ordinary least squares (OLS) regression between the logarithms of errors 
and the logarithms of time steps. 
The mean parallel velocity and energy was evaluated in samples as 
the first and the second order moment function of solutions for investigation.
For both of the two physical quantity, the second-order method is 
near 2.0 higher than the order of the Euler-Maruyama and 
the Cadjan-Ivanov methods which are both near 1.0.
Meanwhile, to achieve the same error level of $\exp(-9)$, 
the second order use the least time nearly half of 
the Cadjan-Ivanov method and 1/20 of the Euler-Maruyama method.

To demonstrate the usage of the second-order method with extended Lorentz operator, 
a practical case of backward runaway phenomenon is simulated. 
Although this is a runaway phenomenon\cite{Dreicer1959,Wesson2011}, 
the electrons initially travel in the direction identical to a DC electric field. 
The backward runaway probability is the probability that such 
runaway occurs at a given initial electron velocity. 
Using the second-order algorithm, the runaway phenomenon was simulated and 
visualized at different times and the runaway probability was calculated. 
In addition, when the initial velocity ran over the $v_{||}-v_\perp$ velocity space, 
the contour of the backward runaway probability was plotted.

The remainder of this paper is organized as follows.
In Section II, we derive the stochastic differential equation governing the electron collision effect in plasmas. 
In Section III, a second-order algorithm under a weak convergence condition is presented and compared with 
the Euler-Maruyama and the Cadjan-Ivanov method.
Finally, in Section IV, the backward runaway probability is calculated using the second order method.

\section{Collision Model of Electrons}

The evolution of the electron distribution is governed by 
a Boltzmann equation with a collision operator. 
We used $f$ to denote the distribution function of electrons. 
The collision operator on the right-hand side of the equation 
encapsulates the collision effects as a partial derivative of 
the distribution function. 
Eq.~ \eqref{eqn:Boltzmann} presents the dominant 
Boltzmann equation of the electron distribution function, 
including the collision effects:
\begin{equation}
    \frac{\partial f}{\partial t}+\bm{v}\cdot \frac{\partial f}{\partial \bm{x}}+\frac{q}{m}(\bm{E}+\bm{v}\times \bm{B})\cdot \frac{\partial f}{\partial \bm{v}}={\left(\frac{\partial f}{\partial t}\right)}_{\mathrm{coll}}.
    \label{eqn:Boltzmann}
\end{equation}

Among the varieties of collision operators, 
a suitable choice for high-energy electrons in plasmas is 
the Lorentz collision operator\cite{helander2005collisional}, which is expressed as follows:
\begin{equation}
    {\left(\frac{\partial f}{\partial t}\right)}_L=\frac{Z_i\mathrm{\Gamma }}{2v^3}\frac{\partial }{\partial \mu }\left(1-{\mu }^2\right)\frac{\partial f}{\partial \mu }\,
    \label{eqn:lorentz} 
\end{equation}
where $\mu =v_{||}/v$ represents the velocity parallel to electric field, 
$Z_i$ is the charge carried by ions, and $\mathrm{\Gamma }$ is a constant 
defined by $\mathrm{\Gamma }=n_eq^4\mathrm{ln}\mathrm{\Lambda }/4\pi {\epsilon }^2_0m^2_e$, 
indicating the collision intensity. This operator reflects the collision effects of electrons 
with ions under two assumptions. 
The first assumption states that the ions are uniformly distributed in the background, 
whereas the second states that the mass of ions is much greater than that of electrons.

To incorporate the effects of collisions between the energetic and background electrons 
and ions under the velocity limit condition $v\gg v_T$ for the extended Lorentz operator, 
a velocity friction term is added\cite{Karney1986,shkarofsky1992numerical}, which is expressed as follows:
\begin{equation}
    {\left(\frac{\partial f}{\partial t}\right)}_{friction}=\mathrm{\Gamma }\frac{\bm{v}}{v^3}\cdot \frac{\partial f}{\partial \bm{v}}.
    \label{eqn:friction}
\end{equation}
In addition, the original pitch-angle-scattering effect caused by the ion-electron collision
\begin{equation}
    \left(\frac{\partial f}{\partial t}\right)_{scatter}=\frac{\mathrm{\Gamma Z_i}}{2v^3}\frac{\partial }{\partial \mu }\left(1-{\mu }^2\right)\frac{\partial f}{\partial \mu }
    \label{eqn:scatter_ei}
\end{equation}
was enhanced by the electron-electron collision 
\begin{equation}
    \left(\frac{\partial f}{\partial t}\right)_{scatter}=\frac{\mathrm{\Gamma }}{2v^3}\frac{\partial }{\partial \mu }\left(1-{\mu }^2\right)\frac{\partial f}{\partial \mu }.
    \label{eqn:scatter_ee}
\end{equation}
The extended Lorentz collision operator including both effects for energetic electrons can be expressed as follows:
\begin{equation}
    {\left(\frac{\partial f}{\partial t}\right)}_{\mathrm{extended}}=\mathrm{\Gamma }\frac{\bm{v}}{v^3}\cdot \frac{\partial f}{\partial \bm{v}}+\frac{\mathrm{\Gamma }\left(1+Z_i\right)}{2v^3}\frac{\partial }{\partial \mu }\left(1-{\mu }^2\right)\frac{\partial f}{\partial \mu }.
    \label{eqn:colltotal}
\end{equation}

To solve the Boltzmann equation Eq.~\eqref{eqn:Boltzmann} with the extended 
Lorentz collision operator Eq.~\eqref{eqn:colltotal}, 
we determined an SDE using the Dynkin formula\cite{Oeksendal2003,Klebaner2005} 
on the collisional Boltzmann equation. We initiated from the Boltzmann equation 
for energetic electrons in Cartesian coordinates, i.e., 
\begin{equation}
 \frac{\partial f}{\partial t}+\left(\frac{q\bm{E}}{m}+\frac{q}{m}\bm{v}\times \bm{B}-\mathrm{\Gamma }\frac{\bm{v}}{v^3}\right)\cdot \frac{\partial f}{\partial \bm{v}}+\frac{{(Z}_i\mathrm{+1)}\mathrm{\Gamma }}{2}\frac{\partial }{\partial \bm{v}}\cdot \left(\frac{v^2I-\bm{vv}}{v^3}\cdot \frac{\partial f}{\partial \bm{v}}\right)=0.
    \label{eqn:cartcoor}
\end{equation}
The corresponding Ito SDE has the following general form:
\begin{equation}
  d\bm{v}\left(t\right)=\bm{\mu }\left(\bm{v},t\right)dt+\sum^m_{j=1}{{\bm{\sigma }}_j\left(\bm{v},t\right)dW^j}.
    \label{eqn:generalSDE} 
\end{equation}
Required the distribution of the solution of Eq.~\eqref{eqn:generalSDE} exactly solves 
the original Boltzmann equation Eq.~\eqref{eqn:Boltzmann}, 
the coefficient of Eq.~\eqref{eqn:generalSDE} should satisfy the following relations:
\begin{align}
    \bm{\sigma }{\bm{\sigma }}^T&=\left(Z_i+1\right)\mathrm{\Gamma }\cdot \frac{v^2\bm{I}-\bm{vv}}{v^3}, \label{eqn:condsigma} \\
    \bm{\mu }&=\frac{q\bm{E}}{m}+\frac{q}{m}\bm{v}\times \bm{B}-(2+Z_i)\mathit{\Gamma}\frac{\bm{v}}{v^3}. \label{eqn:condmu}
\end{align}
According to Dynkin formula, the distribution of solution 
of such an Ito SDE solves the Boltzmann equation. 
It can be directly shown that the right-hand side of Eq.~\eqref{eqn:condsigma} 
comprises the entries in a positive definite matrix. Furthermore, the $\bm{\sigma} $ matrix is 
the square root of this matrix. Therefore, $\bm{\sigma}$ 
can be numerically solved via Cholesky decomposition. 
By further demanding for $\bm{\sigma} $ to be symmetric, $\bm{\sigma}$ 
can be solved in a closed form using Cadjan and Ivanov's decomposition 
method\cite{Cadjan1999,Cadjan1997} as follows:
\begin{equation}
  \bm{\sigma }=\sqrt{\frac{\left(Z_i+1\right)\mathrm{\Gamma }}{v}}\left(\bm{I}-\frac{\bm{vv}}{v^2}\right).
    \label{eqn:sigmacloseform}
\end{equation}
Plugging the Eq.~\eqref{eqn:condmu} and Eq.~\eqref{eqn:sigmacloseform} into 
Eq.~\eqref{eqn:generalSDE}, the extended Ito SDE can be rewritten in vector form as follows:
\begin{equation}
  d\bm{v}=\left(\frac{q}{m}\bm{E}\bm{+}\frac{q}{m}\bm{v}\times \bm{B}-(2+Z_i)\mathrm{\Gamma }\frac{\bm{v}}{v^3}\right)dt-\sqrt{\frac{{(Z}_i+1)\mathrm{\Gamma }}{v^5}}\bm{v}\times \bm{v}\times d\bm{W}. 
    \label{eqn:totalsde} 
\end{equation}

The left-hand side of this SDE is the infinitesimal increment of the velocity process, 
whereas the right-hand side is the sum of the two terms. 
The first term is the total force experienced by a particle. 
It contains a Lorentz force generated by an external field 
and the friction arising from collisions. 
The second term is the effect of pitch-angle scattering. 
This term corresponds to the random force in the Langevin equation; 
however, it is written here in a handy notation that 
allows it be understood as an Ito integral in the Wiener process.

\section{Higher-order weak convergence algorithms}

Traditional measurement of an algorithm for solving an SDE in Monte Carlo 
simulation for collisions is focus on the path accuracy, 
which can be expressed by the definition of the strong convergence condition. 
To express it formally, let $X_T$ be the theoretical solution and 
$X^{\Delta }$ be the numerical solution. 
If there exist constants $C$ and ${\Delta }_0$ independent of the time step $\Delta $, 
for any $\Delta \in (0,{\Delta }_0)$,
\begin{equation}
    E\left[\left|X_T-X^{\Delta }(T)\right|\right]\le C{\Delta }^{\gamma },
    \label{eqn:strong_conv}
\end{equation}
we can say that $X^{\Delta }$ is strongly convergent\cite{kloeden1992higher, Kloeden1992} 
to $X_T$ with order $\gamma $.

While, for Monte Carlo simulation of collisions, 
it is the distribution function or its moments rather than the sample path itself matters. 
Therefore, the measurement of a numerical algorithm should be established based on 
the difference between the distributions, which is the weak convergence condition. 
Formally, this condition can be expressed as follows:
\begin{equation}
    \left|E\left[g(X_T)\right]-E\left[g(X^{\Delta}(T))\right]\right|\le C{\Delta }^{\beta },
    \label{eqn:weak_conv}
\end{equation}
where $g\in C^{2\left(\beta +1\right)}\left({\mathcal{R}}^d,\mathcal{R}\right)$ is any continuous function. 
If a numerical solution, $X^{\Delta }$, satisfies Eq.~\eqref{eqn:weak_conv}, 
it is weakly convergent\cite{Kloeden1992} to $X$ at time $T$ with order $\beta $. 
Some simple choices of function $g$ help make this definition more intuitive. 
If random process $X(t)$ is taken as the velocity process and $g(X)=X$, 
weak convergence gives the expectation convergence condition, which implies the correct mean velocity. 
If $g(X)=(X-\bar{X})^2$, this then yields the variance convergence condition, which implies the correct temperature. 
For all continuous $g$, if Eq.~\eqref{eqn:weak_conv} hold, which guarantees all order of the moments.

The strong convergence condition sufficient guarantees the week convergence condition\cite{Kloeden1992}. 
However, to meet the week convergence condition, 
the strong condition is not necessarily to be satisfied. 
What is worse, the improvement of strong convergence order 
does not sufficiently results in the improvement of week convergence order. 
The Milstein method\cite{Kloeden1992}, for example, have the first order strong convergence
higher than the Euler-Maruyama which has the 1/2 order strong convergence. 
But they have the same the first order of weak convergence. 
Therefore, this kind of accuracy increment of sample path with 
additional computation cost is in vain for the distribution accuracy improvement. 
As a consequence, instead of improving the accuracy of electron trajectories, 
we directly seek more accurate distribution functions. 
Thus, a higher-order weak convergence algorithm is employed.

A second-order weak convergence numerical method is constructed as follows. 
To find the weak solution of the SDE with the general form 
\begin{equation}
  dX^k={\mu }^k\left(t,X\right)dt+\sum^m_{j=1}{{\sigma }^{k,j}\left(t,X\right)dW_j},
\end{equation}
a function of its solution can be expanded into multiple Ito integrals\cite{Klebaner2005,Kloeden1992} as follows:
\begin{equation}
f\left(X_{\tau }\right)=\sum_{\alpha\in\mathcal{A}} I_\alpha{\left[f_\alpha\left(\rho ,X_{\rho}\right)\right]}_{\rho ,\tau }+\sum_{\alpha\in\mathcal{B}(\mathcal{A})} I_\alpha{\left[f_\alpha\left(\cdot ,X_{\cdot }\right)\right]}_{\rho ,\tau },
\label{eqn:Ito_multiples}
\end{equation}
where $\mathcal{A}$ is a hierarchical set, $\mathcal{B}(\mathcal{A})$ are 
the corresponding remainder sets\cite{Kloeden1992}, 
and $\rho$ and $\tau$ are two stopping times that indicating the start and end time points. 
This is a stochastic analog of the deterministic Taylor expansion. 
The first term on the right-hand side of Eq.~\eqref{eqn:Ito_multiples} 
is the expansion series to a certain order, and the second term is the remainder term. 
By dropping the remainder term, an approximation of function $f$ is obtained. 

Because the weak convergence condition is used, the simplification can go further. 
Each Ito integral can be estimated using the following equation\cite{Kloeden1992}:
\begin{equation*}
I_{\alpha }{\left[f_{\alpha }(t_0,X_{t_0})\right]}_{t_0,t}\approx f_{\alpha }\left(t_0,X_{t_0}\right){\hat{I}}_{\alpha ,t_0,t}.
\end{equation*}
Further substitution of multiple Ito integrals must satisfy the convergence condition. 
As for second-order weak convergence, the condition can be simplified to 
\begin{equation}
\left|E[\hat{I}]\right|+\left|E[{\hat{I}}^2]-\mathrm{\Delta }\right|+\left|E[{\hat{I}}^3]\right|+\left|E[{\hat{I}}^4]-3{\mathrm{\Delta }}^2\right|+\left|E[{\hat{I}}^5]\right|\le K{\mathrm{\Delta }}^3.
\end{equation}
We introduce a simple three-point distribution of random variables\cite{Kloeden1992} 
$\mathrm{\Delta }{\hat{W}}^j$ as $\hat{I}$, which is defined as follows:
\begin{equation}
  P\left(\mathrm{\Delta }{\hat{W}}^j=\pm \sqrt{3\mathrm{\Delta }}\right)=\frac{1}{6},\qquad P\left(\mathrm{\Delta }{\hat{W}}^j=0\right)=\frac{2}{3}.  \label{eqn:dwhat}.
\end{equation}
It can be verified that in both the three-point and classic Gaussian $N(0;\mathrm{\Delta })$ distributions, 
the random variables meet the moment conditions. 
However, it is much easier and cheaper to generate and compute $\Delta\hat{W}$. 
This simplifies the calculation while maintaining the order of weak convergence.

An explicit second-order weak convergence algorithm is then constructed as follows\cite{Kloeden1992}. 
We define two supporting vector variables as follows:
\begin{align*}
    {\overline{R}}^j_{\pm } &=X_n+\mu \left(X_n\right)\mathrm{\Delta }\pm {\sigma }^j(X_n)\sqrt{\mathrm{\Delta }},\\
    {\overline{U}}^j_{\pm } &=X_n\pm {\sigma }^j\left(X_n\right)\sqrt{\mathrm{\Delta }}.
\end{align*}
Using these variables, numerical multiple integrals of the first and second order are constructed as follows:
\begin{equation*}
    \mathrm{\Upsilon}_{c1}=\frac{1}{4}\sum_{j=1}^m\left[\sigma^j(\overline{R}^j_+)+\sigma^j(\overline{R}^j_-)+2\sigma^j(Y_n)+\sum_{ \substack{ r=1\\ r\neq j}}^m
\left(\sigma^j(\overline{U}^r_+)+\sigma^j(U^r_-)-2\sigma^j(Y_n)\right)\right]\mathrm{\Delta}{\hat{W}^j},
\end{equation*}
\begin{equation*}
    \mathrm{\Upsilon}_{c2}=\frac{{\mathrm{\Delta}}^{-\frac{1}{2}}}{4}\sum_{j=1}^m\left[\left(\sigma^j(\overline{R}^j_+)-\sigma^j(\overline{R}^j_-)\right)\left(\left(\mathrm{\Delta}{\hat{W}}^j\right)^2-\mathrm{\Delta }\right)+\sum^m_{\substack{r=1\\ r\neq j}}
\left(\sigma^j(\overline{U}^r_+)-\sigma^j(\overline{U}^r_-)\right)(\mathrm{\Delta }\hat{W}^j\mathrm{\Delta }\hat{W}^r+V_{r, j})\right],
\end{equation*}
where
\begin{equation*}
V=\begin{pmatrix}
-\mathrm{\Delta } & \cdots  & {\xi }_{j_1,{\ j}_m} \\
\vdots  & \ddots  & \vdots  \\
{-\xi }_{j_1,{\ j}_m} & \cdots  & -\mathrm{\Delta } \end{pmatrix}
\end{equation*}
is a stochastic matrix with entries ${\xi }_{ij}$ set to be 
random variables taking the value $\Delta$ or $-\Delta$ with the same probability. 

Finally, the second-order algorithm can be expressed as follows:
\begin{align}
    \overline{\mathrm{\Upsilon }}&=Y_n+\mu \left(Y_n\right)\mathrm{\Delta }+\sum^m_{j=1}{b^j\mathrm{\Delta }{\hat{W}}^j}, \label{eqn:finalalg1} \\
    Y_{n+1}&=Y_n+\frac{1}{2}\left(a\left(\overline{\mathrm{\Upsilon }}\right)+a\left(Y_n\right)\right)\mathrm{\Delta }+{\mathrm{\Upsilon }}_{c1}+{\mathrm{\Upsilon }}_{c2}. \label{eqn:finalalg2}
\end{align}

The first two terms on the right-hand side of Eq.~\eqref{eqn:finalalg2} correspond to 
the deterministic Euler method using a predictor-corrector scheme. 
The remaining two terms are the modifications for diffusion. 
${\mathrm{\Upsilon }}_{c1}$ yields the first-order integral term of the Ito expansion, 
and ${\mathrm{\Upsilon }}_{c2}$ improves the accuracy of the final results up to the second order. 

\section{Backward runaway probability}
\subsection{Phenomenon and theory}
The runaway effect, in a uniform electric field in space and without a magnetic field, 
results from the competition between collisional friction and external electric forces. 
A collision comprises two sources\cite{Karney1986}: the effect of energetic electrons colliding with 
heavy and static ions and the effect of energetic electrons colliding with background thermal electrons. 
Both kinds of collision will affect pitch-angle scattering. 
But only the collision with background electrons yields a frictional force on energetic electrons. 
If the electric force is greater than the collisional friction, the electron will be further accelerated. 
Higher velocity reduces the friction, hence the acceleration continues. 
This is called the runaway phenomenon.
On the contrary, if the collisional friction is greater than the electric force, 
the electron will be slowed down. Lower velocity increases the friction until the 
velocity is reduced to the thermal velocity at the end. This is call being stopped.

The backward runaway mechanism is similar to that for standard forward runaways 
but with its initial velocity direction opposite to that of the electric force. 
Because the initial velocity moves in opposition to the electric force, 
the electron is pulled to slow down by the electric force at the beginning. 
When the speed is reduced to near the thermal velocity of electrons $v_T$, 
the high-velocity approximation is violated and the collision operator is no longer valid\cite{Fisch1987}. 
The collision frequency is assumed to be constant\cite{Fisch1987} there and 
the electrons are subsequently thermalized to be normally distributed in velocity space 
as background electrons. 
But if an electron gains a sufficient perpendicular velocity from 
pitch-angle scattering\cite{Fernandez2012, Fisch1987}, it can avoid this stopping phenomenon. 
Moreover, if the electric force exerted on the unstopped electron overcomes the dynamic friction, 
the electron can be indefinitely accelerated as forward runaway. 
This is called the backward runaway. Because the collision force of the scattering effect is a random effect, 
there is an uncertainty whether an electron will be stopped or classified as runaway. 
This uncertainty is measured by the backward runaway probability, 
which can be estimated by the fraction of electrons that are not stopped out of the total number of electrons.

According to the dynamics of backward runaway, 
the collision operator for energetic electrons in Eq.~\eqref{eqn:colltotal}, 
which includes collision effects from both background ions and electrons, is used. 
By introducing the characteristic variables, 
the dominate equation Eq.~\ref{eqn:totalsde} is simplified into a dimensionless form. 
We introduce the Dreicer velocity\cite{Dreicer1959} $v_D$ and 
the characteristic collision time $\tau =v^3_D/\mathrm{\Gamma }$. 
The renormalized variables are $\tilde{v}=v/v_D$, ~$\tilde{t}=t/\tau $, $\tilde{E}=qE/F_D$, 
and~$\tilde{W}=W(\tau \tilde{t})/\sqrt{\tau }\ $, which is a standard Wiener process as $W(t)$. 
The critical force is defined as $F_D=\mathrm{\Gamma }m_e/v^3_d$. 
Then, the dimensionless SDE can be written as follows:
\begin{equation}
  d\tilde{v}=\left(\tilde{E}-\left(2+Z_i\right)\frac{\tilde{v}}{{\tilde{v}}^3}\right)d\tilde{t}+{\sqrt{1+Z_i}\tilde{v}}^{-5/2}\tilde{v}\times \tilde{v}\times d\tilde{W}.
    \label{eqn:dimless}
\end{equation}

When the velocity of an energetic electron is less than the Dreicer velocity, 
it is labeled as a stopped electron. 
As a result, Eq.~\eqref{eqn:dimless} is no longer valid and the velocity of 
the stopped electron is simulated by sampling from the thermal distribution of background electrons. 
Furthermore, we did not consider the secondary electron emission problem, 
i.e., once an energetic electron is stopped, 
there is no way for it to become excited again.

At the end of simulation at time $T$, 
the runaway probability $P_r$ can be defined as follows:
\begin{equation}
    P_r=E\left[I_r(X_T)\right],
    \label{eqn:def_runaway_prob}
\end{equation}
where $I_r(X)$ is an indicator function defined as follows:
\begin{equation}
    I_r(X)=
    \begin{cases}
    1 & X\ge v_D, \\
    0 & otherwise.
    \end{cases}
    \label{eqn:indicator_func}
\end{equation}

\subsection{Estimation of the order of algorithms}
The weak order of the algorithm is defined according to Eq.~\eqref{eqn:weak_conv}. 
We estimate the numerical order of the second method, the Euler-Maruyama and 
the Cadjan-Ivanov method in the backward runaway simulation case. 
The numerical orders of the three methods are compared.
The numerical order is calculated according to the following procedure.
First, given the time step and numerically solve the SDE Eq.~\eqref{eqn:dimless}.
By averaging the function $g(X^\Delta_T)$ to estimate the expectation $E[g(X^\Delta_T)]$.
Second, change the step size and calculate the expectation $E[g(X^\Delta_T)]$ with the same way.
Third, perform the ordinary least square (OLS) on the logarithms of 
the error of expectation $E[g(X^\Delta_T)]$ and the logarithms of the time steps $\Delta_t$
to obtain the slope, which is the order estimated.

The third in the procedure above need to know the theoretical solution to calculate errors in practice,
but the theoretical solution of equation eq.~\eqref{eqn:dimless} is unknown.
To overcome this difficulty, we find out a good enough solution to replace the theoretical solution.
The solution is so accurate that first, it cannot be distinguished from that of nearest larger time steps and 
second, it cannot be distinguished from the solution of different method at the same time step.

To test the first requirement, since the variances of solution differ in different time steps,
the Welch's unequal variance t-test is used to detect the difference of two solutions.
If the null hypothesis of the test cannot be rejected, we cannot tell the difference and 
thus it satisfies the first requirement.
In addition, Shapiro test is performed to check the normality of solution distribution
in order to meet the presumption of Welch's t-test.

To test the second requirement, the analysis of variance (ANOVA) technique is used.
The ANOVA F-test is utilized to detect the means of different method is equal or not.
If the solution is normally distributed and the F-test fails to reject the null hypothesis,
the second requirement is satisfied.

Given an SDE, the estimation of $E[g(X^\Delta_T)]$ and corresponding variance is calculated
according to the following algorithm.

\begin{algorithm}[H]
    \caption{Estimate $E[g(X^\Delta_T)]$ and corresponding variance}
    \KwData{Sample number in a batch $N$, total batch number $M$, continuous function $g(x)$, end time $T$, time step $\Delta$}
    \KwResult{$E[g(X^\Delta_T)]$}
    \For{$m \gets0$ \KwTo $M-1$}{
        \quad \For{$n \gets0$ \KwTo $N-1$}{
        \qquad Generate a sample of Wiener process $W$\;
        \qquad Solve the SDE by Wiener sample to obtain the numerical solution $X_T^\Delta(n)$\;
        }
        \quad calculate the batch mean $E[g(X^\Delta_T)](m) = \sum_n g[X^\Delta_T(n)]/N$ on $N$ sample in one batch\;
    }
    calculate the mean and variance of $E[g(X^\Delta_T)]$ of M batches\;
\end{algorithm}

In the simulation, we set $N = 100,000$ samples in each batch and $M = 30$ total batches. 
The initial time step is $\Delta\tilde{t}=1.0$.
The simulation scans from $K=0$ where $\Delta\tilde{t}=2^{0}$ to $K=6$ 
where $\Delta\tilde{t}=2^{-6}$ in total 7 different steps.
All simulations end at $\tilde{t}=1.0$. 
In addition, $Z_i=1$ and the electric field is $\tilde{E}=\{1,0,0\}$, 
and the initial velocity is $\tilde{v} = \{3,0,0\}$. 

\begin{figure}
    \centering
    \includegraphics[width=0.8\linewidth]{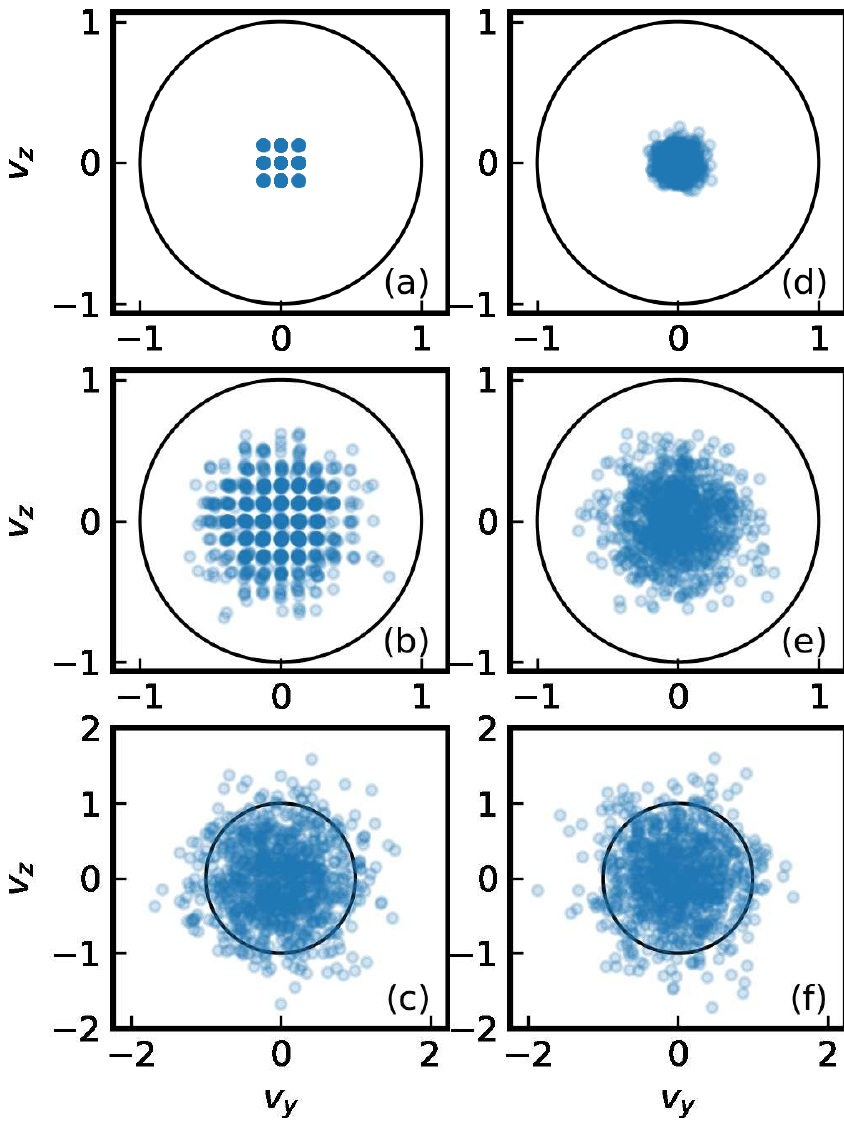}
    \caption{The samples of step size being $2^{-7}$ obtained from 
    the second order method at time step 1, 10, 50 are plotted in 
    the left column (a), (b) and (c). And the results of the Euler-Maruyama method at 
    corresponding time step 1, 10, 50 are plotted in 
    the right column (d), (e) and (f).\label{fig:weak_property}}
\end{figure}

The samples of solution at different time step using the Euler-Maruyama method 
and the weak order method are plotted in the Fig.~\ref{fig:weak_property}. 
Noticed that the second order method only guarantees the accuracy of the distribution function and 
the path is simplified using a three point distribution, Eq.~\eqref{eqn:dwhat}, 
the solution from the second order method is first spitted into grids but keep the moment correct. 
With collision accumulation, the distribution obtained from different method converges.

To demonstrate the numerical order of the algorithm, 
we choose two typical functions with clear physical meanings. 
Noticed average of the $y$ and $z$ components of the solution are 
both 0 due to the system's symmetry, 
our first function is chosen to be the $x$ velocity component, i.e., 
$g_1(X)=X_x$, which is a first-order moment function. 
The second is the energy function $g_2(X)=X_x^2+X_y^2+X_z^2$, 
which is a second-order moment function. 

\begin{figure}
    \centering
    \includegraphics[width=0.8\linewidth]{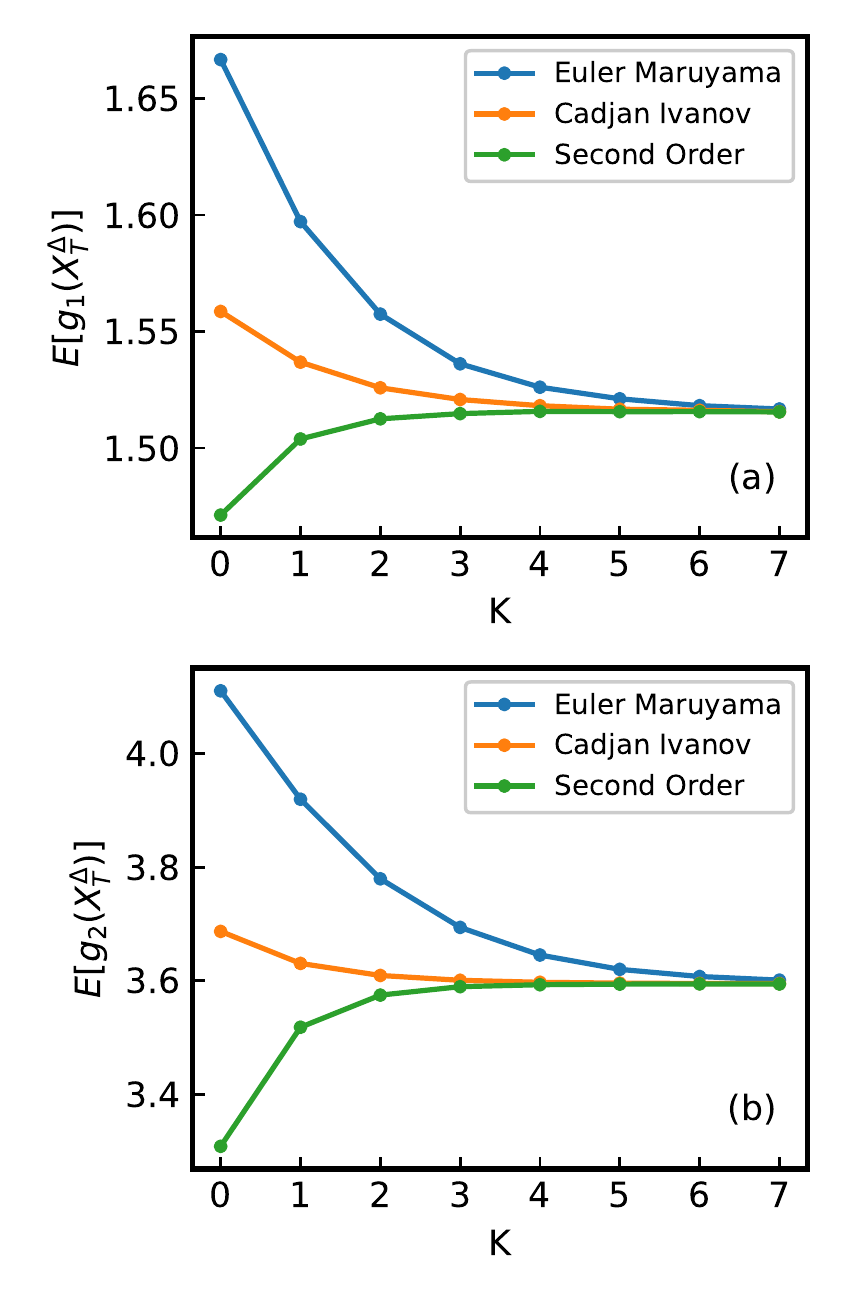}
    \caption{The first-order moment functions $E[g_1(X^\Delta)]$s using the Cadjan-Ivanov method (blue), 
    the Euler-Maruyama method (orange), and the second-order method (green) are plotted in (a). 
    The second-order moment functions $E[g_2(X^\Delta)]$ using 
    the three methods are plotted in (b).\label{fig:moments_values}}
\end{figure}

The expectation of $g_1$ and $g_2$ with respect to time step is plotted in 
Fig.(\ref{fig:moments_values}). All three numerical methods converge to 
the same value with decreasing time steps as $K$ increasing. 
The first moments, $E[g_1(X^\Delta)]$ are plotted in 
Panel (a) using the three algorithms and the second moments, 
$E[g_2(X^\Delta)]$, are plotted in Panel (b). 

As discussed above, we next choose a good enough solution as the theoretical solution that 
satisfied the two requirements.
First, solutions with the same method in two different time steps is investigated.
The solution of $K=7$ is selected as potential target and is compared with the solution of $K=6$.
To quantitatively demonstrate that the two solutions under $K=5$ and $K=6$ are sufficiently close, 
we perform Welch's unequal variance t-tests for the functions of the first- and 
second-order moments. 
The normality of different method in $K=5$ and $K=6$ are tested using Shapiro test, 
and the statistics are listed in table.~\ref{tab:shapiro_test}
Under Welch's unequal variance t-test on $E[g_1(X^\Delta)]$ and $E[g_2(X^\Delta)]$, 
We calculate the statistics ($t$), degrees of freedom ($\nu$), 
and corresponding p-values in table.~\ref{tab:ttest_K}.

From table.~\ref{tab:ttest_K}, when the time step is reduced from $K=6$ to $K=7$,
the $p$-value of second order method on two moments function are all above 50\%.
We cannot reject the null hypothesis that the two solution are the same, so the 
solution of second order method with $K=7$ passes the first test.

\begin{sidewaystable}
    \scriptsize
    \centering
    \caption{Shapiro normality tests of the moment functions, $E[g_1(X_T^\Delta)]$ and $E[g_2(X_T^\Delta)]$,
     of the Euler-Maruyama, the Cadjan-Ivanov method and the second-order method 
    at time step of $K=6$ and $K=7$.
    \label{tab:shapiro_test}}
    \begin{tabular}{|c|c |c|c|c|c |c|c|c |c|c|c|}
        \hline
        Expectation                         & Methods        & \multicolumn{2}{|c|}{$K=6$} & \multicolumn{2}{|c|}{$K=7$} \\
        \cline{3-6}
                                            &                & W-statistic & p-value        & W-statistic & p-value       \\
        \hline
        \multirow{3}*{$E[g_1(X_T^\Delta)]$} & Euler-Maruyama &  0.9288509488 & 0.0457747914 & 0.9448584914 &  0.1229689941 \\
                                            & Cadjan-Inanov  &  0.9723019600 & 0.6038923264 & 0.9418374896 &  0.1019422486 \\
                                            & Second-Order   &  0.9591118097 &  0.2939443886 & 0.9595355392 &  0.3014061153 \\
        \hline
        \multirow{3}*{$E[g_2(X_T^\Delta)]$} & Euler-Maruyama & 0.9620262384 & 0.3486383259  &  0.9357385635 & 0.0698712692 \\
                                            & Cadjan-Inanov  & 0.9853706956 & 0.9433261156  &  0.9483121037 & 0.1523194462 \\
                                            & Second-Order   & 0.9637868404 & 0.3855869174  &  0.9749118686 & 0.6801327467 \\
        \hline
    \end{tabular}
    
    \bigskip\bigskip  
    
    \caption{Welch's unequal variance tests of the moment functions, $E[g_1(X_T^\Delta)]$ and $E[g_2(X_T^\Delta)]$,
    at two different time step of $K=6$ and $K=7$
    using the Euler-Maruyama, the Cadjan-Ivanov method and the second-order method.
    statistics $t$, degree of freedom $\nu$, and the $p$-value are listed in the table.
    The $p$-values of the Second-order method are all above 50\%.
    \label{tab:ttest_K}}
      \begin{tabular}{|c|c |c|c|c|c |c|c|c |c|c|c|}
        \hline
        Expection & Methods & \multicolumn{3}{|c|}{$K=6$}& \multicolumn{3}{|c|}{$K=7$} & t & $\nu$ &  p-value \\
        \cline{3-8} 
                  &         &    M &   Mean & Std        & M & Mean & Std              &    &       & (two-sided)\\ 
        \hline
        \multirow{3}*{$E[g_1(X_T^\Delta)]$} & Euler-Maruyama & 30 & 1.5181471154 & 0.0010572291 & 30 & 1.5167099144 & 0.0009388375 & -5.4738626822  & 57.2007133978 & 0.0000010189 \\
                                            & Cadjan-Inanov  & 30 & 1.5161602266 & 0.0011350644 & 30 & 1.5155411902 & 0.0008255942 & -2.3751147326  & 52.9744270096 & 0.0211956264 \\
                                            & Second-Order   & 30 & 1.5155835766 & 0.0011201136 & 30 & 1.5154497894 &  0.0011551838 & -0.4477535215 & 57.9449611347 & 0.6560006370 \\
        \hline
        \multirow{3}*{$E[g_2(X_T^\Delta)]$} & Euler-Maruyama & 30 & 3.6075405807 & 0.0011471987 & 30 & 3.6011830706 & 0.0007982998 &-24.4960634146 & 51.7508692636 & 0.0000000000 \\
                                            & Cadjan-Inanov  & 30 & 3.5952511325 & 0.0005412185 & 30 & 3.5945317579 & 0.0007720412 & -4.1554688610 & 52.4304111506 & 0.0001202918 \\
                                            & Second-Order   & 30 & 3.5944935081 & 0.0008078345 & 30 & 3.5943761437 & 0.0008143028 & -0.5510099036 & 57.9963114240 & 0.5837434234 \\
        \hline
        \end{tabular}
    
\end{sidewaystable}
    
\begin{sidewaystable}
    \scriptsize
    \centering
    \caption{ANOVA table of the moment functions, $E[g_1(X_T^\Delta)]$ and $E[g_2(X_T^\Delta)]$,
    at two different time step $K=7$ using the Cadjan-Ivanov method and 
    the second-order method with and without the Euler-Maruyama method. \label{tab:ttest_methods}}
    \begin{tabular}{|c|c |c|c|c|c |c|c|}
        \hline
        Expectation & Methods Set & & Sum Squares & Degree of Freedom & Mean Squares & F-statistic & p-value \\ 
        \hline
        \multirow{2}*{$E[g_1(X^\Delta)]$} & Euler-Maruyama and & Treatments &   0.0000296219 &    2 &   0.0000148109 &  14.8238525409 & 0.0000028839 \\
                                          &  Cadjan-Ivanov and & Error      &   0.0000869241 &   87 &   0.0000009991 &                &              \\
                                          &       Second Order & Total      &   0.0001165460 &   89 &                &                &              \\
        \hline
        \multirow{2}*{$E[g_1(X^\Delta)]$} & Cadjan-Ivanov and  & Treatments &   0.0000001253 &    1 &   0.0000001253 &   0.1201698569 & 0.7301048951 \\
                                          &      Second-Order  & Error      &   0.0000604817 &   58 &   0.0000010428 &                &              \\
                                          &                    & Total      &   0.0000606070 &   59 &                &                &              \\
        \hline
        \multirow{2}*{$E[g_2(X^\Delta)]$} & Euler-Maruyama and & Treatments &   0.0009059843 &    2 &   0.0004529922 & 700.0487590997 & 0.0000000000 \\
                                          &  Cadjan-Ivanov and & Error      &   0.0000562965 &   87 &   0.0000006471 &                &              \\
                                          &       Second-Order & Total      &   0.0009622808 &   89 &                &                &              \\
        \hline
        \multirow{2}*{$E[g_2(X^\Delta)]$} & Cadjan-Ivanov and  & Treatments &   0.0000003632 &    1 &   0.0000003632 &  0.5666709067  & 0.4546294465 \\
                                          &      Second-Order  & Error      &   0.0000371781 &   58 &   0.0000006410 &                &              \\
                                          &                    & Total      &   0.0000375413 &   59 &                &                &              \\
        \hline
    \end{tabular}
    
    \bigskip\bigskip  
    
    \caption{Ordinary Least Squares regression over logarithms of numerical errors and logarithms of time steps
        using the Euler-Maruyama, the Cadjan-Ivanov and the second Order method
        on $E[g_1(X^\Delta)]$ and $E[g_2(X^\Delta)]$ respectively. Slopes of the Euler-Maruyama and the Cadjan-Inanov
        are near 1 and the slope of the second order method is near 2.
        \label{tab:slopes}}
    \begin{tabular}{|c|c |c|c|c|c|c|}
        \hline
        Expectation & Method & slope mean & slope std. & slope t-statistics & p-value & $R^2$ \\
        \hline
        \multirow{3}*{$E[g_1(X^\Delta)]$} & Euler-Maruyama & -0.9833 &  0.011 & -93.483 & 0.000 & 0.999  \\
                                          & Cadjan-Ivanov  & -1.0038 &  0.018 & -56.783 & 0.000 & 0.998  \\
                                          & Second-order   & -1.7603 &  0.087 & -20.265 & 0.000 & 0.990  \\
        \hline
        \multirow{3}*{$E[g_2(X^\Delta)]$} & Euler-Maruyama & -0.9091 &  0.020 & -45.462 & 0.000 & 0.997  \\
                                          & Cadjan-Ivanov  & -1.1313 &  0.042 & -27.137 & 0.000 & 0.993  \\
                                          & Second-order   & -2.0024 &  0.027 & -75.199 & 0.000 & 0.999  \\
        \hline
    \end{tabular}
\end{sidewaystable}

For the second test, given the fixed time step to be $K=7$, the equality of 
mean of solutions from the different methods are tested using AVOVA technique. 
Since the normality has been verified using shapiro test before in table.~\ref{tab:shapiro_test}, 
we directly conduct the AVOVA.
We perform the AVOVA on two groups of methods, one consists of 
all three methods and the other group only consists of the Cadjan-Ivanov and the second order method.
The analysis results are listed in table.~\ref{tab:ttest_methods}.

According to table.~\ref{tab:ttest_methods},
in the group of all three methods, the means
are significantly different regardless of 
$E[g_1(X_T^\Delta)]$ or $E[g_2(X_T^\Delta)]$, 
the main error is bought by the Euler-Maruyama method.
But in the group containing only the Cadjan-Inanov and the second order method,
we cannot reject the null hypothesis of the means are the same.
Therefore, we select the solution from the second order method under 
$K=7$ as the accurate solution and 
used to compute error of other methods and other time steps.

\begin{figure}
    \centering
    \includegraphics[width=0.8\linewidth]{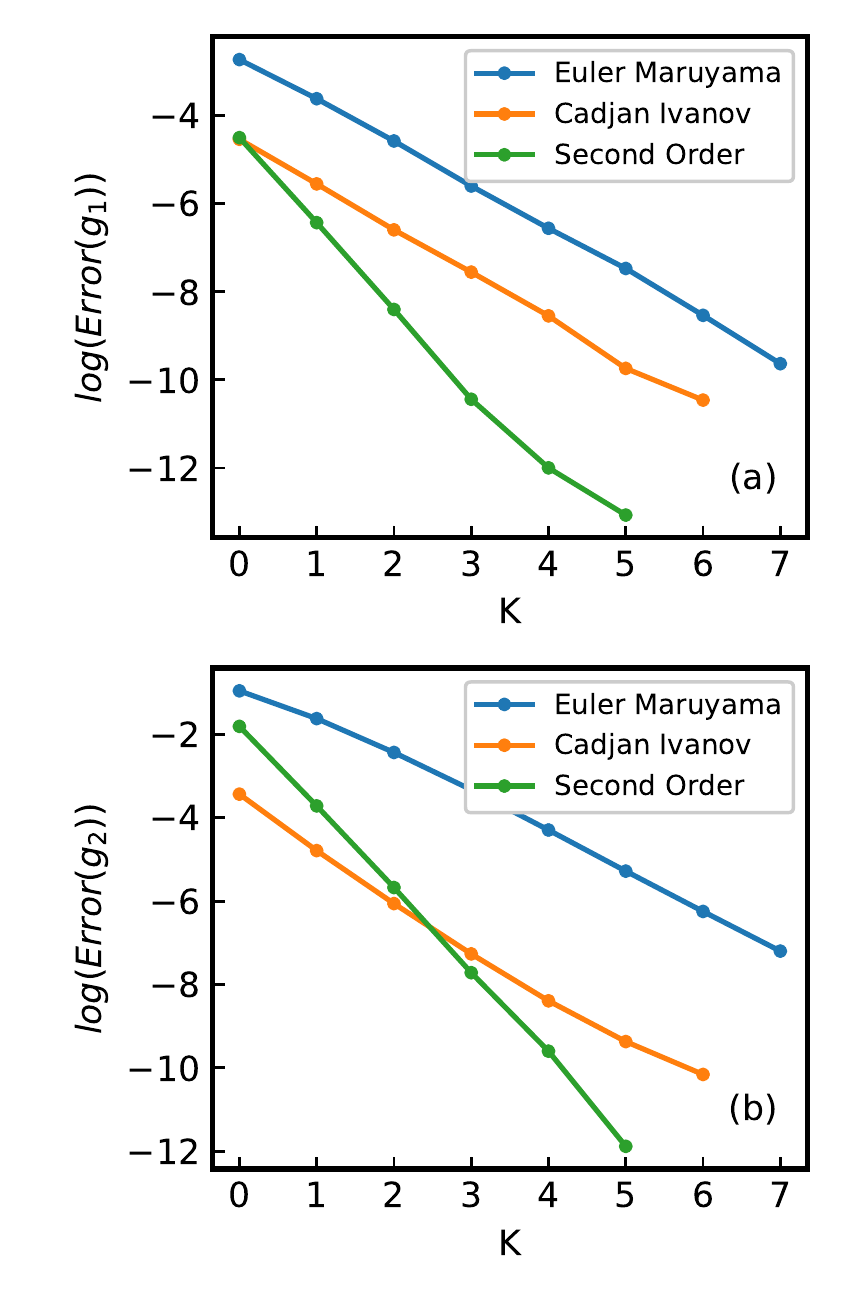}
    \caption{Logarithms of errors of $E[g_1(X^\Delta)]$ (a) and $E[g_2(X^\Delta)]$ (b)
    at different logarithms of time steps $K$ using 
    the Euler-Maruyama method (blue)，the Cadjan-Ivanov (orange) 
    and the second-order methods (green).\label{fig:logerr}}
\end{figure}

The error is calculated and the numerical order of 
algorithm is obtained from OLS regression as slope
according to the logarithms of errors and logarithms of time steps. 
The result is plotted in Fig.~\ref{fig:logerr} and statistics are shown in table.~\ref{tab:slopes}

The errors of the first order moments $E[g_1(X_T^\Delta)]$ and 
second order moments $E[g_2(X_T^\Delta)]$ are plotted in subplots 
Fig.~\ref{fig:logerr}a and Fig.~\ref{fig:logerr}b, respectively. 
We can observe that the errors decrease when the time step is decreased for all algorithms. 
The order of the algorithm is estimated by the slope in table.~\ref{tab:slopes}.
The Euler-Maruyama and the Cadjan-Ivanov methods have slope of 0.98 and 1.00 respectively
for $E[g_1(X_T^\Delta)]$. And the second order has a slope of 1.76.
For $E[g_2(X_T^\Delta)]$, the Euler-Maruyama and the Cadjan-Ivanov have the slope of 0.92 and 1.21.
The second order method has a slope of 2.00.
As expected, the second order method has a slope near 2.0 higher than these of the Euler-Maruyama and
the Cadjan-Ivanov method near 1.0 for both functions.
Even the other method may have a smaller error at large time steps, with the time step decreasing,
the second order can reduce the error much faster.
After $K>3$, the second order method perform better than all other methods.

In the aspect of time consumption, to reduce the error of
 $E[g_1(X_T^\Delta)]$ to the level of $\exp(-9)$,
it costs the Euler-Maruyama method 26.374s
and the Cadjan-Ivanov method 3.411s. But the
the second order method only uses 1.510s.
To reduce the error of $E[g_2(X_T^\Delta)]$ to the same level,
it costs the Euler-Maruyama method 95.94s 
and the Cadjan-Ivanov method 6.796s. The
the second order method uses 5.780s.
The higher requirement of accuracy, the less time the second order method uses
compared to the Euler-Maruyama method and the Cadjan-Ivanov method.

\subsection{Runaway probability contour}

The fixed ion background is set to contain protons that have $Z_i=1$. 
The electric field is uniformly distributed in the space and taken to be $\tilde{E}=(1,0,0)$ 
and no magnetic field is set. The initial velocities of the electrons 
are $\tilde{v}_{||}=10$ and $\tilde{v}_{\bot}=0$. 
Considering the time step to be $\mathrm{\Delta }\tilde{t}=0.01$, 
the simulation ends at $\tilde{t}=30$.

\begin{figure}
    \centering
    \includegraphics[width = 0.8\linewidth]{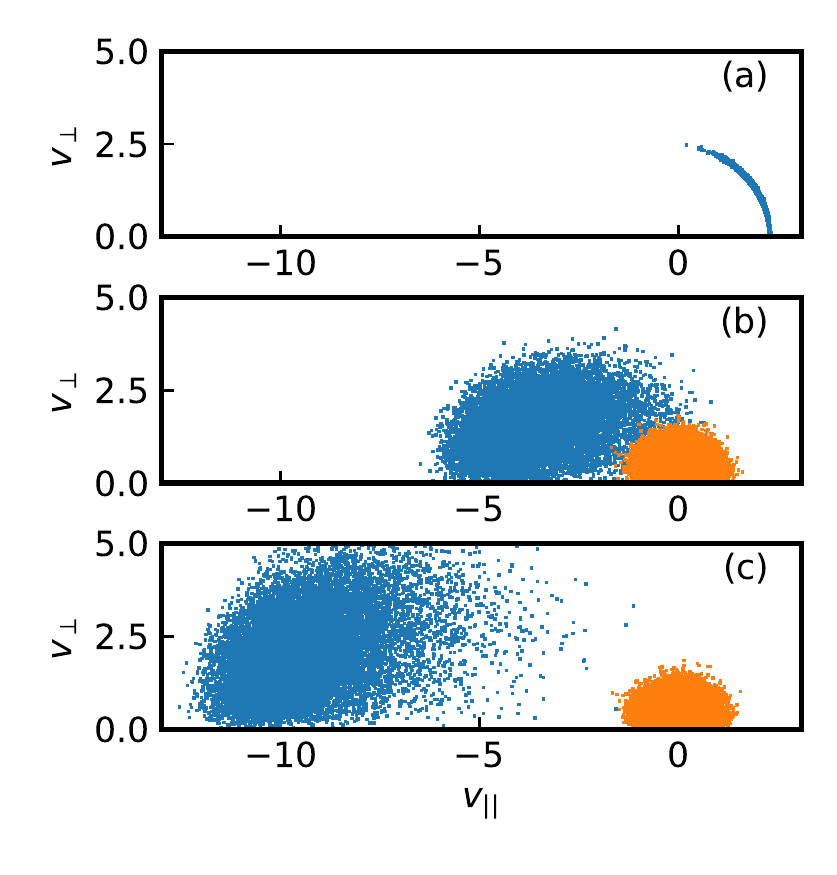}
    \caption{Scattering plot of 10,000 backward runaway electron samples at different times. 
    All samples started with $\tilde{v}_{||} = 3$ and $\tilde{v}_\bot = 0$. 
    Under the collision effects, some samples achieved sufficient perpendicular velocity to 
    runaway (blue), while others were stopped (red).\label{fig:runaway_sim}}
\end{figure}

We solve the SDE Eq.~\eqref{eqn:dimless} to obtain 30 batches of solutions, 
and each batch contains 10,000 samples. 
One of the typical batches on different time is plotted in Fig.~\ref{fig:runaway_sim}. 
The velocities are plotted in velocity space at three different time points: 
$\tilde{t}=6,\ \tilde{t}=14$, and $\tilde{t}=20$. 
The horizontal and vertical axes denote the parallel velocity $\tilde{v}_{||}$ and 
the vertical velocity $\tilde{v}_{\bot }$, respectively. 
Red samples are those whose velocity is less than $v_D$. 
They are thermally distributed as background electrons. 
On the contrary, blue points are samples that are considered to be eventual runaways.

The transient runaway probability time is calculated as the fraction of unstopped electron at a given time. 
Using time step size on $K=5$, we calculate the transient runaway probability of electron starting from different initial velocity.
\begin{figure}
    \centering
    \includegraphics[width = 0.8\linewidth]{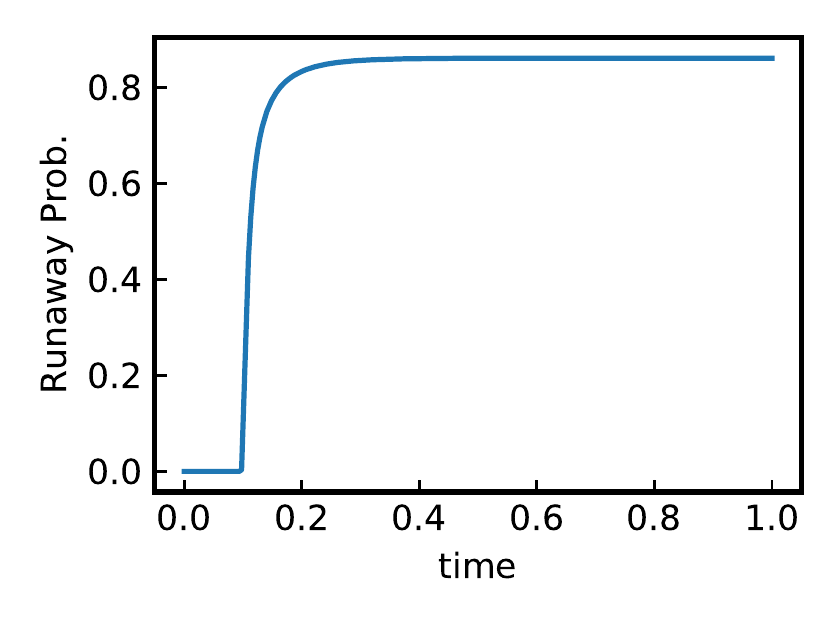}
    \caption{Transient runaway probability at different initial parallel velocity\label{fig:trans_runaway_prob}}
\end{figure}

From Fig.~\ref{fig:trans_runaway_prob}, we verified that the t end to be 30 is a valid choice.

The runaway probability in Eq.~\eqref{eqn:def_runaway_prob} in expectation form can be approximated by 
the average of function $I_r(X_T)$. 
Noticed that the indicator function take the value of one only when the sample is runaway, 
this expression is equivalent to the fraction of runaway sample to the total sample number. 
By calculating the ratio of runaway, the runaway probability is obtained. 

\begin{figure}
    \centering
    \includegraphics[width=0.8\linewidth]{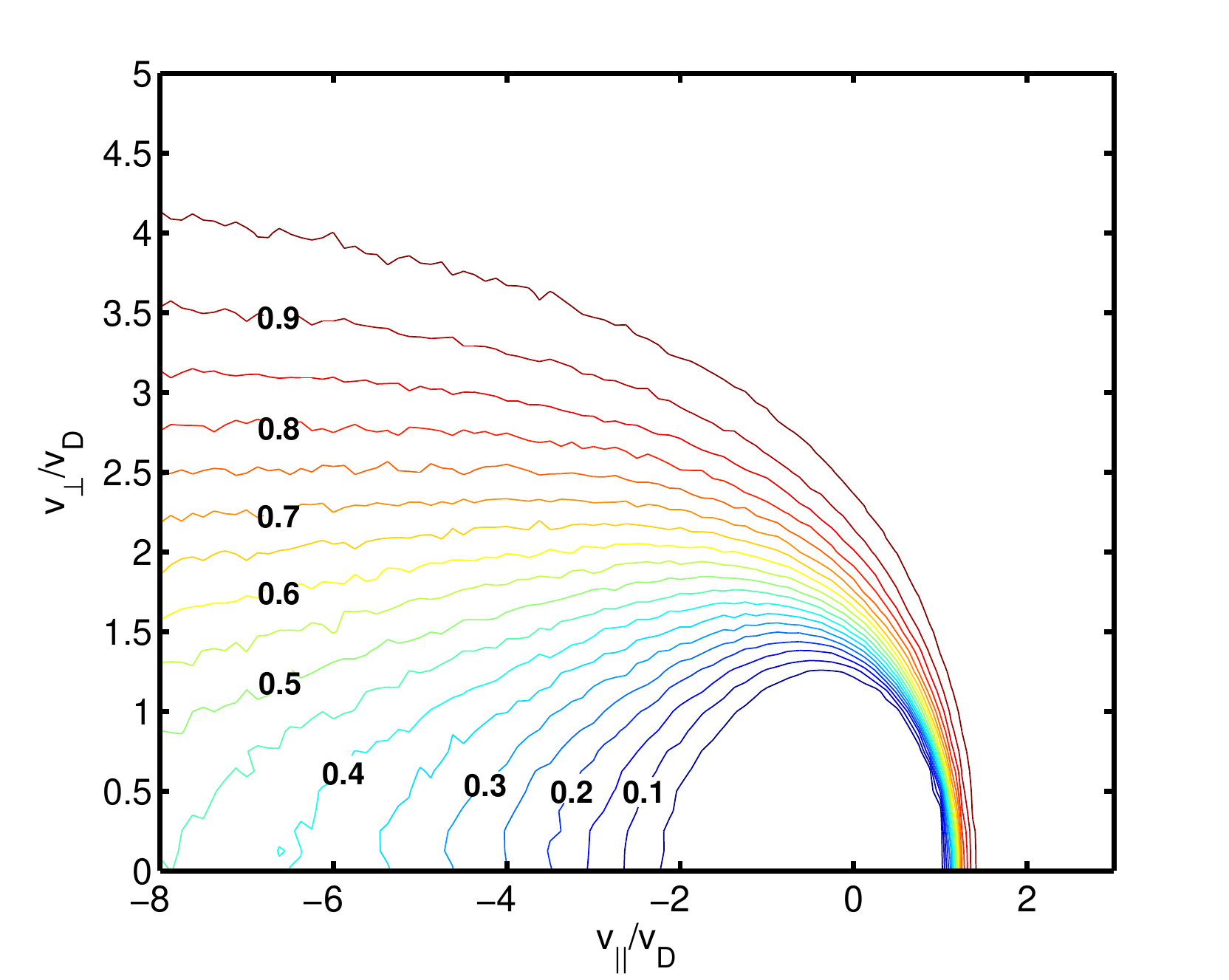}
    \caption{The contour plot of backward runaway probability in velocity space obtained from 
    a Monte Carlo simulation of 3,000 samples. The time step is 0.002, 
    and the time ends at $\tilde{t}=20$. 
    Nuclear charge is set to $Z_i=1$.\label{fig:runaway_prob}}
\end{figure}

Next, we calculated the backward runaway probability for electrons initiated under different conditions. 
We use $N=3,000$ samples and simulate one batch for each initial velocity. The time step is set to 0.01, 
and the time ends at $\tilde{t}=20$ and $Z_i=1$. The parallel velocity $v_{||}$ varies from $-8.0$ to $3.0$, 
whereas the perpendicular velocity ranges from $0.0$ to $5.0$. The contour is plotted in Fig.~\ref{fig:runaway_prob}. 

It is evident from the figure that for $v_{||}$, 
the backward runaway probability increases as $v_{\bot }$ grows and that for a fixed $v_{\bot }$, 
this probability increases as $v_{||}$ moves away from 0. 
Electrons with $v_{||}$ greater than $v_D$ runaway almost certainly, 
allowing the forward runaway case to be recovered.

\section{Discussion}
Starting with the Boltzmann equation of energetic electrons colliding with static heavy ions and 
background thermal electrons, we developed an SDE for the energetic electrons under both collision effects. 
Instead of aiming to improve the accuracy of the path of the sample, 
we rather aimed to improve the accuracy of the distribution. 
A second-order algorithm was developed to solve the SDE under the weak convergence condition, 
improving the accuracy with which the moments and distributions were calculated. 
The numerical weak order was estimated in the backward-runaway case 
and compared with that estimated using the widely applied the Euler-Maruyama and the Cadjan-Ivanov methods. 
The runaway probability was defined and shown to converge with order same as that of a weak-order algorithm. 
A backward runaway case was simulated, and the contours of runaway probability were calculated.

In future research, structure-preserving algorithms, such as symplectic algorithms\cite{Milstein2002a} 
or volume-preserving algorithms\cite{Qin2013}, could be developed to solve the SDE. 
In addition, a system containing an electron source can be solved using the Feynman-Kac formula, 
after which the collision problem with the source term can thus be solved. 
The energetic electron simulation program may also inspire many applications in geophysics and space physics. 
The advantage of adopting the weak convergence condition is that it improves 
the calculation accuracy of the physical quantities associated 
with distributed computation and specific moment calculations.

\begin{acknowledgments}
acknowledgments
\end{acknowledgments}

\bibliographystyle{aipnum4-1}
\bibliography{stochastic}

\end{document}